# 200Gb/s VCSEL transmission using 60m OM4 MMF and KP4 FEC for AI computing clusters

Tom Wettlin[1], Youxi Lin[1], Nebojsa Stojanovic[1], Stefano Calabrò[1], Ruoxu Wang[2],
Lewei Zhang[2], Maxim Kuschnerov[1]

[1]*Huawei Technologies Düsseldorf GmbH, Riesstr. 25, 80992 Munich, Germany*
[2]*Huawei Technologies, 430000, Wuhan P.R.China*
tom.jonas.wettlin@huawei.com

**Abstract:** We show a beyond 200Gb/s VCSEL transmission experiment. Results are based on 35GHz VCSEL and advanced DSP. We show an AIR of 245Gb/s PAM-6 back-to-back, and 200Gb/s PAM-4 over 60m OM4 fiber assuming KP4-FEC. © 2024 The Author(s)

## 1. Introduction

The market introduction of the public user application ChatGPT led to an unprecedented interest in AI-based computing for large language models, video generation or robot intelligence. Cloud operators are moving towards interconnecting GPU accelerators and high bandwidth memory directly with optics in an effort to scale out the clusters and reduce latency, which requires 3-4 times more bandwidth than pure Ethernet connectivity. As the size of the computing clusters continues to grow, the optical connectivity shifts from shorter active optical cables to vertical-cavity surface-emitting laser (VCSEL) based transceivers with 50-100m reach and parallel single mode fiber optics with 500m link budget. 100Gb/s per lane connectivity dominates the current transceiver generation with native 100Gb/s in the electrical and optical domain. Here, 400G (4x100Gb/s) and 800G (8x100Gb/s) are predominantly deployed. In 2023, the first transceiver prototypes using 200Gb/s per lane optics and 100Gb/s SerDes were announced, which were exclusively DR8 optics based on parallel single mode fiber. Low cost VCSEL based optics so far haven't been demonstrated for rates at 200Gb/s per lane while complying with low latency standard KP4 forward error correction (FEC) and PAM-4 modulation as defined in IEEE 802.3dj Ethernet task force. Previous record demonstrations were using unrealistic assumptions for data center optics, such as digital multi-tone (DMT) and a high complexity and latency soft decision FEC for a net rate of 187Gb/s [1] or high latency FEC for a net rate of 187Gb/s [2]. The latest publication by Broadcom showed a PAM-4 transmitter at 100Gbaud [3], which didn't include any FEC overhead to enable error free transmission at 200Gb/s. Further recent results are listed in [4-6].

In this publication, we demonstrate the VCSEL transmission at rates exceeding 200Gb/s over standard OM4 fiber complying with IEEE standards of PAM-4 and KP4 FEC.

## 2. VCSEL and system characterization

The experimental investigations were done based on a discrete measurement setup. The analog electrical signal is generated by a digital-to-analog converter (DAC) with a 6dB bandwidth of 50GHz at one sample per symbol. The signal is amplified by a SHF M827 B driver amplifier with a 3dB bandwidth of 70GHz, a bias current is added and the signal is directly modulated by an 850nm packaged VCSEL with a 3dB bandwidth of 35GHz and a relative intensity noise (RIN) spectrum as depicted in Figure 1. As it can be computed from Figure 2 the RMS spectrum of the VCSEL is 0.178nm. The output signal is either directly inserted into the detector by a short connection fiber, or transmitted over up to 60m of OM4 multimode fiber (MMF). Signal detection and analog-to-digital conversion (ADC) are done using a Keysight DCA N1092E. Limited by the discrete components and the RF connectors, the overall end-to-end system bandwidth is 28GHz, as shown on Figure 3 for different VCSEL bias currents. Digital signal processing is applied at the transmitter and receiver and described in the following.

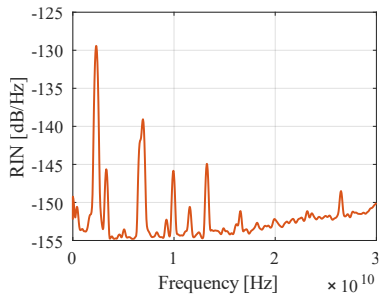 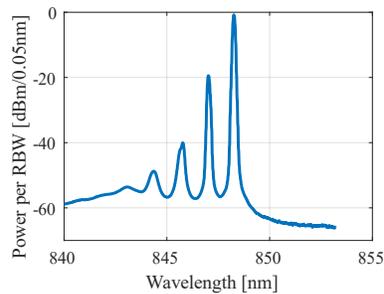 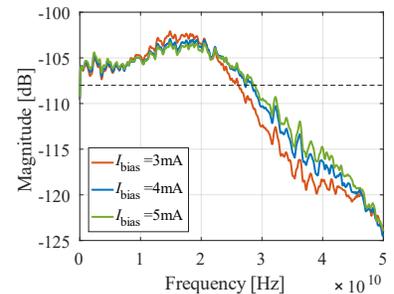

Figure 1 – Received RIN spectrum    Figure 2 – VCSEL optical spectrum    Figure 3 - End-to-end transmission spectrum of the system

## 3. Digital Signal Processing Stack

To cope with the nonlinearities induced by the VCSEL as well as the severe bandwidth limitations of the transmitter and receiver components, multiple DSP schemes are combined. The applied steps are depicted in Figure 4. At the transmitter, the signal is duobinary (DB) pre-coded, pre-skewed by a level dependent filter and a simple digital pre-distortion (DPD) for bandwidth limitations based on a 3-tap finite impulse response (FIR) filter is performed (response shown in Figure 4 inset b)). Note that we use the term *duobinary* also for the equivalent coding schemes for higher order PAM formats. At the receiver, timing recovery is applied, before a chain of simplified Volterra nonlinear equalization (VNLE) targeting the DB response, feed-forward noise cancellation (NC) [7] and maximum likelihood sequence estimation (MLSE) is used to cope with inter-symbol interference (ISI) and nonlinearities.

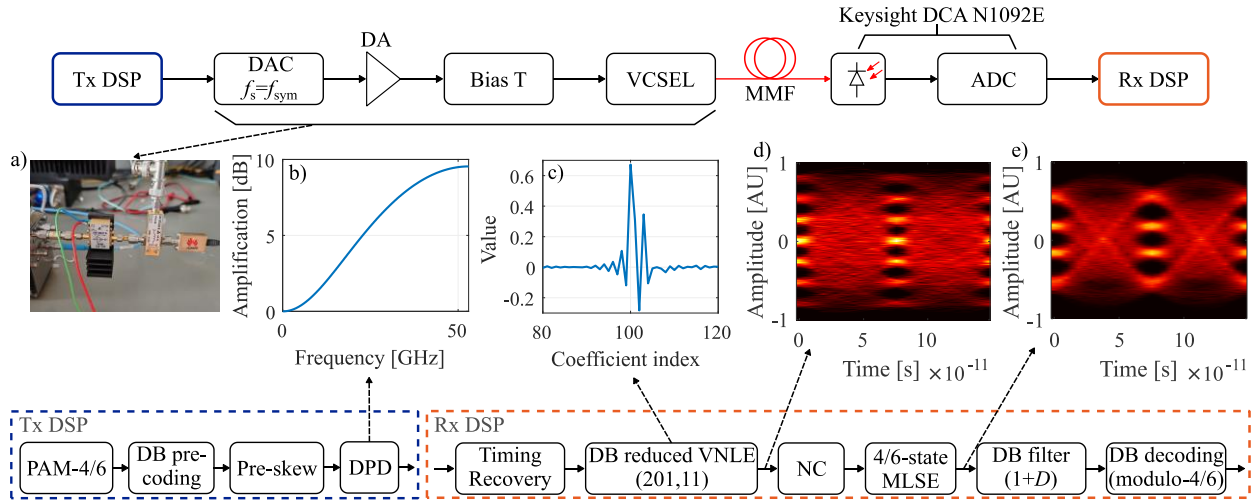

Figure 4 – System and DSP block diagram. Inset a) shows the discrete transmitter setup, b) the preemphasis by the simple DPD, c) the central linear VNLE coefficients, d) the DB PAM-4 eye diagram at 106Gbaud and e) the reconstructed eye diagram after MLSE for 106Gbaud PAM-4.

The number of linear taps in the VNLE was optimized to be 201. This large number is necessary to cope with reflections in the discrete setup and can be significantly reduced in an integrated system. In inset c) of Figure 4 the central 40 linear VNLE coefficients are shown for 106Gbaud PAM-4. As visible, only few central coefficients show a significant value. The nonlinear part of the VNLE is restricted to the main and first super diagonal of the second order terms. A memory length of 11 is chosen, leading to 21 taps for the nonlinearity compensation. Due to the DB $(1 + D)$ target response, the VNLE output has seven levels in case of PAM-4 and eleven levels in case of PAM-6. An eye diagram of 106Gbaud DB PAM-4 is shown in inset d) of Figure 4. For an additional performance improvement, feed-forward NC can be used. This approach is based on estimating the noise as the difference between samples after feed-forward equalization (FFE) and symbol decisions, whitening it and adding it back to the symbol decisions [7]. For the whitening FIR filter, 6 coefficients (3 pre-cursor and 3 post-cursor), obtained by Burg's method [8], are used. To remove the controlled ISI induced by the DB target of the FFE, $M$-state MLSE is applied, resulting in PAM-$M$ decisions. A reconstructed eye diagram of 106Gbaud PAM-4 after MLSE based on the histogram reconstruction technique shown in [9], is depicted in inset e) of Figure 4. Since DB pre-coding is applied at the transmitter, the DB encoding step is applied on the resulting signal, before it is decoded by a modulo-$M$ operation. Note that DB pre-coding can be avoided in case MLSE is applied after the DB FFE, however, it breaks error bursts originating from the quasi catastrophic behavior of the trellis in channels with strong ISI [10].

## 4. Experimental Results

The results for PAM-4 and PAM-6 transmission with different gross rates in a back-to-back configuration are shown in Figure 5. The received optical modulation amplitude (OMA) is 1dBm. Different receiver DSP configurations are considered, applying either a subgroup or all of the approaches described in the previous part. It is visible for both, PAM-4 and PAM-6, that MLSE can improve the performance compared to DB VNLE only. The combination of NC and MLSE brings a significant additional performance improvement. PAM-4 with advanced DSP reaches below the KP4 FEC threshold of $2.2 \cdot 10^{-4}$ at 106Gbaud, yielding a net rate of 200Gb/s. Due to the strong bandwidth limitations of the system, PAM-6 generally shows better performance than PAM-4 at the same bitrate. 96Gbaud PAM-6 allows a net rate of ~226Gb/s using KP4 FEC.

For a general insight about the reachable rates independent from specific codes, the achievable information rate (AIR) considering hard-decision FEC is shown in the inset of Figure 5. For the shown curves, the full DSP stack is applied. PAM-4 reaches a maximum rate of 223Gb/s at 112Gbaud and PAM-6 245Gb/s at 100Gbaud. Due to the strong bandwidth limitations of the system, a low FEC overhead leads to the best results. The optimal AIR region for PAM-6 between 100Gbaud and 106Gbaud assumes overheads of 2% and 8%, respectively.

Next, different lengths of OM4 fiber are introduced into the link. Figure 6 shows the performance of a gross rate of 212Gb/s PAM-4 as a function of the OMA after transmission over 30m and 60m OM4 MMF. The receiver sensitivity after 30m OMA shows no penalties compared to the back-to-back case (comparing the 1dBm point with Figure 5). However, 60m transmission introduces transmission impairments. In this case, the combination of VNLE, NC and MLSE in the receiver DSP allows to reach a BER below the KP4 FEC threshold. Finally, the transmission of a gross rate of 212Gb/s PAM-6 over 60m and 100m OM4 fiber is shown in Figure 7. After 60m, the KP4 FEC threshold is reached using only DB VNLE in the receiver DSP. Increasing the link length to 100m introduces transmission penalties with the performance slightly above the KP4 limit.

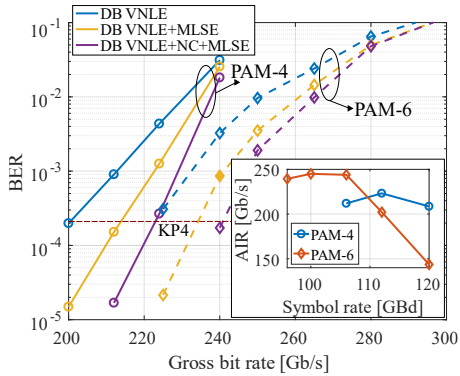
Figure 5 – VCSEL back-to-back performance vs. gross bit rate for PAM-4 and PAM-6

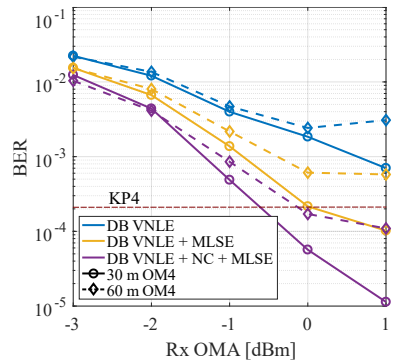
Figure 6 – 212Gb/s PAM-4 transmission over 30m and 60m OM4

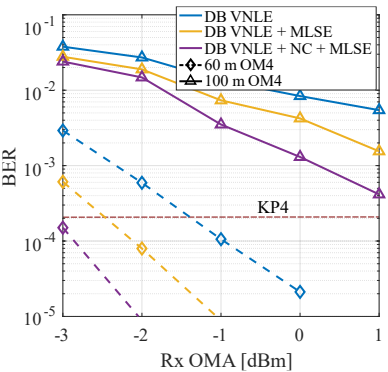
Figure 7 – 212Gb/s PAM-6 transmission over 60m and 100m OM4

## 5. Conclusions

In this publication, net 200Gb/s VCSEL transmission is demonstrated. It utilizes standardized PAM-4 signaling and achieves performance below the Ethernet KP4 FEC for 30m and 60m OM4 MMF transmission. The AIR of the VCSEL reaches as high as 245Gb/s using 100Gbaud PAM-6. This experiment shows the clear feasibility of low cost VCSEL based multi-mode transceivers for 1.6Tb/s Ethernet based on 8x200Gb/s. In the future, we target to further decrease the RMS spectrum of the transmitter, flatten the RIN spectrum and increase the modulation bandwidth, which would enable transmission distances suitable for AI computing cluster pods with limited DSP complexity.